# Integrative Approaches in Cybersecurity, Artificial Intelligence, and Data Management: A Comprehensive Review and Analysis

Marwan Omar[*]
[1] Illinois Institute of Technology
Email: [*] Corresponding momar3@iit.edu

**Abstract**

In recent years, the convergence of cybersecurity, artificial intelligence (AI), and data management has emerged as a critical area of research, driven by the increasing complexity and interdependence of modern technological ecosystems. This paper provides a comprehensive review and analysis of integrative approaches that harness AI techniques to enhance cybersecurity frameworks and optimize data management practices. By exploring the synergies between these domains, we identify key trends, challenges, and future directions that hold the potential to revolutionize the way organizations protect, analyze, and leverage their data. Our findings highlight the necessity of cross-disciplinary strategies that incorporate AI-driven automation, real-time threat detection, and advanced data analytics to build more resilient and adaptive security architectures.

## 1. INTRODUCTION

The rapid advancements in technology have led to the emergence of complex and interconnected systems, driving a significant transformation in the fields of cybersecurity, artificial intelligence (AI), and data management. As organizations continue to rely on vast amounts of data and increasingly sophisticated digital infrastructure, the need for integrative approaches that address the convergence of these domains has become critical. The integration of AI into cybersecurity and data management frameworks offers promising opportunities to enhance security measures, streamline data processes, and mitigate risks associated with cyber threats.

One of the most significant challenges in today's digital landscape is the protection of sensitive information against an ever-evolving array of cyber threats. The integration of AI in cybersecurity has demonstrated considerable potential in addressing these challenges by enabling real-time threat detection, predictive analysis, and automated responses (Zangana, 2024). AI-driven cybersecurity solutions, such as machine learning algorithms and neural networks, can identify patterns and anomalies within vast datasets, allowing for more proactive and adaptive security measures (Omar & Zangana, 2024).

In addition to enhancing cybersecurity, AI's role in data management has become increasingly prominent. The ability to analyze large datasets efficiently and accurately is crucial for organizations to make informed decisions and optimize their operations. AI techniques, such as data mining, natural language processing, and predictive analytics, have been widely adopted to improve data quality, manage information systems, and support decision-making processes (Zangana & Al-Shaikhli, 2013; Zangana, 2018). Furthermore, AI's integration with data management systems has facilitated the development of more robust and scalable solutions, capable of handling the growing complexity of data generated by modern digital ecosystems (Zangana & Abdulazeez, 2023).

The convergence of cybersecurity, AI, and data management also presents significant opportunities for innovation in areas such as distributed computing and edge computing. These advancements have enabled the creation of more resilient and efficient systems, capable of processing and analyzing data closer to its source, thereby reducing latency and improving security (Zangana, Mohammed, & Zeebaree, 2024). Moreover, the adoption of AI-driven automation in data management processes has the potential to significantly reduce the burden on human operators, allowing for more efficient and error-free operations (Zangana, Mohammed, & Mustafa, 2024).

As the digital landscape continues to evolve, the importance of integrative approaches that combine cybersecurity, AI, and data management will only grow. This comprehensive review and analysis aims to explore the synergies between these domains, identify key trends and challenges, and highlight the potential future directions that could shape the next generation of technological solutions. By examining the current state of research and practice in these areas, this paper seeks to contribute to the ongoing discourse on the development of more secure, intelligent, and data-driven systems.

## 2. REVIEW OF AI AND MACHINE LEARNING APPLICATIONS

Artificial Intelligence (AI) and Machine Learning (ML) have revolutionized numerous fields, ranging from computer vision to cybersecurity. The application of AI and ML in these domains has been pivotal in solving complex problems and improving system efficiency. This section provides a comprehensive review of AI and ML applications across various disciplines, highlighting key contributions and advancements.

In the field of computer vision, AI and ML have enabled significant progress in object detection and movement tracking. Al-Sanjary et al. (2018) proposed an innovative approach for detecting object movement using optical flow techniques. This method leverages AI to accurately track and identify moving objects, making it particularly useful in surveillance and automated monitoring systems.

Machine Learning has also played a critical role in enhancing the performance of wireless communication systems. Specifically, in Mobile Ad Hoc Networks (MANETs), hybrid routing protocols that integrate AI techniques have been developed to optimize network performance. Al-Sanjary et al. (2018) conducted an in-depth investigation of these protocols, demonstrating their effectiveness in improving data transmission and reducing latency in dynamic network environments.

Another crucial application of AI is in the domain of social media, where it is used to detect and mitigate the spread of fake news. Khan, Alkawaz, and Zangana (2019) explored the misuse of social media platforms for disseminating false information, emphasizing the role of AI-driven algorithms in identifying and curbing fake news. Their work underscores the importance of AI in maintaining the integrity of online information.

In the realm of data security, AI has been instrumental in developing advanced watermarking techniques for protecting digital content. Majeed (2020) introduced a novel image watermarking method based on the Mojette Transform, which leverages AI to hide and retrieve information securely. This technique is a testament to the potential of AI in enhancing data protection mechanisms.

Moreover, AI's application in shape detection algorithms has contributed significantly to image processing and pattern recognition. Zangana (2017) developed a new algorithm for shape detection that combines multiple AI techniques to achieve high accuracy in identifying geometric shapes. This algorithm has broad implications for fields such as robotics and automated manufacturing, where precise shape recognition is essential.

AI and ML are also being integrated into data management systems to improve data quality and decision-making processes. For instance, Zangana (2018) designed an information management system for pharmacies that utilizes AI to streamline data handling and ensure the accuracy of medical records. Similarly, AI-driven data warehouses have been developed to support student information systems, enabling more efficient management of educational data (Zangana, 2018).

Furthermore, AI has made significant strides in the development of clustering algorithms, which are essential for organizing and analyzing large datasets. Zangana and Abdulazeez (2023) reviewed various clustering algorithms used in engineering applications, highlighting the role of AI in optimizing these techniques for better performance and scalability.

The integration of AI into cybersecurity frameworks has also been extensively explored. Omar and Zangana (2024) edited a comprehensive volume on redefining security with Cyber AI, showcasing how AI-powered applications can enhance security measures, such as network firewall rule analyzers and blockchain-based timestamping tools. These applications demonstrate the transformative impact of AI on cybersecurity, offering new ways to protect digital assets and maintain system integrity.

**Table 1: Summary of AI and ML Applications in Different Domains**

| Domain | Key AI/ML Applications | References |
|---|---|---|
| Computer Vision | Object detection, movement tracking | Al-Sanjary et al. (2018) |
| Wireless Communication | Hybrid routing protocols for MANETs | Al-Sanjary et al. (2018) |
| Social Media | Fake news detection and mitigation | Khan, Alkawaz, and Zangana (2019) |
| Data Security | Watermarking techniques | Majeed (2020) |
| Image Processing | Shape detection algorithms | Zangana (2017) |
| Data Management | Pharmacy information systems, student information systems | Zangana (2018) |
| Clustering Algorithms | Optimizing performance and scalability | Zangana and Abdulazeez (2023) |
| Cybersecurity | Network firewall rule analyzers, blockchain-based timestamping | Omar and Zangana (2024) |

In conclusion, AI and Machine Learning have become indispensable tools in numerous fields, driving innovation and improving efficiency across various applications. From computer vision and wireless communication to data management and cybersecurity, the contributions of AI and ML are vast and continue to evolve, paving the way for future advancements.

## 3. CYBERSECURITY: EMERGING TRENDS AND CHALLENGES

The rapid evolution of technology has introduced significant advancements in cybersecurity, while simultaneously presenting new challenges. One of the most pressing issues in the current digital landscape is the proliferation of sophisticated cyber threats. These threats are increasingly targeting various sectors, exploiting vulnerabilities in systems, and undermining the security posture of organizations (Zangana et al., 2024).

A critical trend in cybersecurity is the growing reliance on Artificial Intelligence (AI) to enhance threat detection and response. AI-driven security solutions are capable of analyzing vast amounts of data, identifying patterns, and predicting potential threats with greater accuracy than traditional methods. This approach is particularly effective in identifying anomalies and preventing security breaches before they occur (Zangana & Zeebaree, 2024). However, as AI becomes more integral to cybersecurity, it also introduces new risks, such as the potential for AI systems to be manipulated or evaded by advanced cyber threats (Omar & Zangana, 2024).

Another emerging challenge is the increasing use of mobile devices, which has expanded the attack surface for cybercriminals. Mobile security threats, such as phishing, malware, and unauthorized access, have become more prevalent, necessitating robust security measures to protect sensitive information stored on mobile platforms (Zangana & Omar, 2020). The integration of mobile devices into enterprise environments further complicates security management, as organizations must ensure that all devices comply with security policies and standards.

The rise of social media as a platform for spreading misinformation and fake news is also a significant concern in cybersecurity. The ability to rapidly disseminate false information can have serious implications for public safety and trust in information systems (Khan et al., 2019). Addressing this issue requires a combination of technological solutions, such as AI-driven content analysis, and public awareness campaigns to educate users about the risks associated with consuming and sharing unverified information.

Moreover, the advent of blockchain technology has been heralded as a potential game-changer in enhancing data security. Blockchain's decentralized and immutable nature makes it a promising tool for securing transactions and data exchanges. However, the implementation of blockchain in cybersecurity is still in its nascent stages, and challenges related to scalability, regulatory compliance, and integration with existing systems must be addressed (Zangana, 2024).

In the realm of network security, the development of advanced firewall rule analyzers has been instrumental in strengthening organizational defenses. These tools help in optimizing firewall configurations, ensuring that security rules are both effective and efficient in mitigating potential threats (Zangana et al., 2024). However, as network infrastructures become more complex, the task of managing and analyzing firewall rules becomes increasingly challenging, necessitating the adoption of more sophisticated tools and techniques.

**Table 2: Key Trends and Challenges in the Convergence of Cybersecurity, AI, and Data Management**

| Trend | Challenge |
|---|---|
| Real-time threat detection and response | Developing accurate and scalable AI-based threat detection models |
| AI-driven automation in data management | Ensuring data integrity and security in automated processes |
| Distributed and edge computing | Balancing performance, security, and resource efficiency |
| Cross-disciplinary integration and collaboration | Fostering interdisciplinary research and knowledge sharing |
| Ethical and regulatory considerations | Addressing privacy concerns and developing governance frameworks |

In conclusion, while significant progress has been made in enhancing cybersecurity, emerging trends and challenges underscore the need for continuous innovation and adaptation. Organizations must remain vigilant and proactive in adopting new technologies and strategies to safeguard against the ever-evolving threat landscape.

**Table 3: Future Directions in the Convergence of Cybersecurity, AI, and Data Management**

| Future Direction | Potential Impact |
|---|---|
| Adaptive and self-learning security architectures | Improved resilience against evolving cyber threats |
| Blockchain-based data management and security | Enhanced data integrity, transparency, and auditability |
| Federated learning and privacy-preserving AI | Collaborative and privacy-sensitive AI models |
| Quantum-resistant cryptographic techniques | Strengthened data protection against quantum computing threats |
| Integrated XR (extended reality) for cybersecurity training | Immersive and experiential security skill development |

## 4. DATA MANAGEMENT AND QUALITY ASSURANCE

Data management and quality assurance are critical components in the information technology domain, influencing the effectiveness of decision-making processes and the accuracy of system outputs. Effective data management practices ensure that data is correctly gathered, stored, and analyzed, while quality assurance processes guarantee the reliability and validity of this data.

A foundational aspect of data management is the establishment of data warehouses, which consolidate data from various sources, providing a unified framework for data analysis. For example, Zangana (2018a) highlighted the development of a data warehouse for a student information system, emphasizing its role in improving data accessibility and analytical capabilities. This approach not only enhances data retrieval efficiency but also supports complex queries, enabling better decision-making.

Quality assurance in data management involves rigorous testing and validation processes to ensure data accuracy and integrity. Zangana (2017a) discussed a case study on library data quality maturity, where systematic evaluation and continuous improvement practices were implemented to enhance data quality. Such practices are vital in ensuring that the data used in various applications remains accurate and reliable over time.

Additionally, the integration of mobile devices into data management systems has introduced new challenges and opportunities. Zangana (2020a) explored the integration of mobile devices in the International Islamic University Malaysia (IIUM) services, which required robust data management protocols to handle the increased data flow and ensure quality across platforms. This integration highlights the need for adaptive data management strategies that can cater to evolving technological landscapes.

In the context of information systems, the design and implementation of management systems are crucial for maintaining data integrity. Zangana (2018b) focused on the design of an information management system for a pharmacy, which underscores the importance of structured data management processes in maintaining operational efficiency and ensuring that critical information is accurately recorded and retrieved.

Furthermore, the accuracy of data is paramount in image processing and analysis. Majeed (2020) introduced a watermarking technique based on the Mojette Transform, demonstrating how data integrity can be preserved in digital images, which is essential in fields where data manipulation or loss can have significant consequences.

The proliferation of social media and the internet has also brought about challenges in managing the quality of information disseminated online. Khan et al. (2019) addressed the misuse of social media for spreading fake news, illustrating the need for stringent data management practices to curb misinformation and protect the integrity of online content.

In conclusion, data management and quality assurance are intertwined processes that are essential for maintaining the accuracy, reliability, and integrity of data across various domains. The continuous evolution of technology necessitates ongoing refinement of these practices to address emerging challenges and ensure that data remains a reliable asset in decision-making processes.

## 5. COMMUNICATION SYSTEMS AND SOCIAL MEDIA

The evolution of communication systems has played a pivotal role in the advancement of social media platforms, enabling rapid information dissemination and interactive connectivity across the globe. Communication systems, such as Mobile Ad Hoc Networks (MANETs), have facilitated the development of robust and flexible networking structures, which are integral to the functioning of modern social media (Al-Sanjary et al., 2018). The characteristics and performance of hybrid routing protocols in MANETs highlight the efficiency and reliability needed to support the high data traffic and dynamic environments characteristic of social media platforms (Al-Sanjary, Ahmed, Zangana, Ali, Aldulaimi, & Alkawaz, 2018).

Social media platforms, while providing unprecedented opportunities for communication and information exchange, have also become channels for the spread of misinformation and fake news. The misuse of social media for such purposes poses significant challenges, as the rapid spread of false information can have widespread consequences. Khan, Alkawaz, and Zangana (2019) explored the use and abuse of social media, emphasizing the importance of developing strategies to mitigate the impact of fake news. Their research underscores the need for more sophisticated detection systems and regulatory frameworks to combat the spread of misinformation on these platforms.

The integration of advanced communication systems and security protocols is essential in addressing these challenges. The deployment of watermarking techniques, as discussed by Majeed (2020), can be a vital tool in protecting the integrity of information shared across social media. By embedding digital watermarks into images and other content, it becomes possible to trace and verify the authenticity of the information, thereby reducing the potential for misuse.

In the context of cybersecurity, communication systems must also be resilient against various threats that target social media platforms. The integration of artificial intelligence (AI) in communication systems can enhance their ability to detect and mitigate these threats. As noted by Omar and Zangana (2024), AI-driven cybersecurity solutions are crucial in identifying and responding to sophisticated cyber threats that exploit social media networks. The fusion of AI with communication systems not only strengthens security but also ensures the reliability of social media as a tool for communication.

In conclusion, the synergy between communication systems and social media is both transformative and challenging. While these technologies have greatly enhanced global connectivity and information exchange, they also require vigilant management and advanced security measures to mitigate the risks associated with their misuse. The ongoing research and development in this field are critical to ensuring that communication systems continue to support the positive potential of social media while safeguarding against its vulnerabilities.

## 6. DECENTRALIZED AND COLLABORATIVE COMPUTING

Decentralized and collaborative computing has emerged as a crucial area in modern computing systems, offering various benefits in terms of flexibility, scalability, and fault tolerance. The shift from centralized to decentralized systems is largely driven by the need for improved data distribution and processing capabilities, particularly in environments where traditional centralized architectures may be limited. These systems leverage multiple distributed nodes to perform computations, store data, and provide services in a manner that enhances reliability and performance.

One of the key applications of decentralized computing is in Mobile Ad Hoc Networks (MANETs), where the absence of a fixed infrastructure necessitates the use of dynamic routing protocols. Al-Sanjary et al. (2018) explored the performance of hybrid routing protocols in MANETs, highlighting their ability to adapt to network topology changes and maintain efficient communication despite the challenges posed by node mobility and variable network conditions. The study emphasizes the importance of decentralized routing mechanisms in ensuring the robustness of MANETs, particularly in scenarios involving high mobility and limited bandwidth.

In the context of edge computing, decentralized and collaborative models are increasingly being integrated into cloud architectures to enhance computing at the network edge. This approach is particularly useful in applications that require low-latency processing and real-time decision-making. Zangana, Mohammed, and Zeebaree (2024) conducted a systematic review of decentralized and collaborative computing models in cloud architectures, focusing on their role in distributed edge computing. The study underscores the potential of these models to reduce latency, improve data locality, and provide better resource utilization in edge environments. This is especially relevant as the demand for real-time data processing continues to grow in various sectors, including healthcare, finance, and smart cities.

Moreover, collaborative computing frameworks are essential in supporting swarm intelligence techniques, which rely on the collective behavior of decentralized agents to solve complex problems. Zangana, Sallow, Alkawaz, and Omar (2024) reviewed the application of swarm intelligence in problem-solving and optimization, demonstrating how decentralized systems can harness the power of collective intelligence to achieve optimal solutions. The study illustrates the effectiveness of collaborative computing in scenarios where traditional centralized approaches may struggle due to the complexity and scale of the problems involved.

Decentralized and collaborative computing also plays a significant role in enhancing network security. By distributing security functions across multiple nodes, these systems can improve resilience against attacks that target centralized points of failure. Omar and Zangana (2024) highlighted the use of decentralized security mechanisms in protecting network infrastructures, particularly in the context of cyber AI. Their work explores the integration of blockchain-based timestamping tools and other decentralized security solutions to create more robust and tamper-resistant systems.

In conclusion, decentralized and collaborative computing represents a paradigm shift that offers numerous advantages in terms of scalability, flexibility, and security. As the demand for real-time processing and resilient computing infrastructures continues to rise, the adoption of these models is likely to expand, driving further innovation in distributed systems and cloud computing.

## 7. CASE STUDIES AND PRACTICAL IMPLEMENTATIONS

The application of theoretical concepts in real-world scenarios is critical for validating their efficacy and understanding their limitations. Case studies and practical implementations offer valuable insights into the challenges and opportunities presented by different technologies and methodologies.

In one study, the detection of object movement using an optical flow approach was explored, demonstrating the practical implications of this method in various industrial applications. The research, conducted by Al-Sanjary et al. (2018), provided a detailed analysis of how optical flow can be effectively used to track and clone object movements in dynamic environments, particularly within computer applications and industrial electronics.

The study of hybrid routing protocols in Mobile Ad Hoc Networks (MANETs) by Al-Sanjary et al. (2018) is another example of a practical implementation that offers significant contributions to the field. The research investigated the characteristics and performance of these protocols, highlighting their potential to enhance communication in decentralized networks. The findings from this study have broad implications for the development of more efficient and reliable MANET systems.

Moreover, the use and abuse of social media for spreading fake news has been critically examined by Khan, Alkawaz, and Zangana (2019). Their research identified the mechanisms through which misinformation spreads on social media platforms and proposed strategies for mitigating this issue. This case study is particularly relevant in the context of cybersecurity, where the rapid dissemination of false information can have far-reaching consequences.

In the domain of image processing, Majeed (2020) explored the use of the Mojette Transform for watermarking images to hide information. This practical implementation highlights the potential of the Mojette Transform in enhancing the security and privacy of digital images, which is increasingly important in the age of pervasive digital communication.

Additionally, the contributions made by Omar and Zangana (2024) in their edited volume on redefining security with cyber AI provide a comprehensive overview of the latest advancements in cybersecurity. The book offers numerous case studies that demonstrate the practical applications of AI in enhancing security measures, thereby bridging the gap between theoretical research and real-world implementation.

Lastly, the development of a new skin color-based face detection algorithm by Zangana (2015) exemplifies the practical application of image processing techniques in security and identification systems. This algorithm, which combines three color model algorithms, has shown to improve the accuracy of face detection systems, making it a valuable tool in various security applications.

These case studies and practical implementations underscore the importance of applying theoretical concepts to real-world scenarios. They not only validate the effectiveness of these concepts but also provide critical insights that can inform future research and development efforts.

## 8. DISCUSSION

The convergence of cybersecurity, artificial intelligence (AI), and data management has led to transformative advancements across multiple industries. This discussion delves into the implications of these integrative approaches, emphasizing the challenges and opportunities presented by their adoption.

The integration of AI in cybersecurity, as explored by Omar and Zangana (2024), illustrates the significant potential of AI-driven tools in enhancing security protocols. AI's ability to analyze vast amounts of data rapidly and accurately makes it indispensable in detecting and mitigating security threats. However, the implementation of AI in cybersecurity is not without challenges. Issues such as algorithmic bias, the need for large datasets, and the complexity of AI models pose significant hurdles. Moreover, the reliance on AI systems introduces new vulnerabilities, as these systems themselves can become targets for sophisticated cyberattacks.

Data management is another critical area where AI is making substantial contributions. As highlighted by Zangana and Omar (2024), AI-powered data management systems offer improved efficiency in handling large datasets, enabling better decision-making and resource allocation. The practical implementations discussed in the case studies demonstrate how AI can optimize data processing and storage, particularly in cloud computing environments. However, the discussion must also acknowledge the ethical and privacy concerns associated with AI-driven data management. Ensuring that data is handled securely and in compliance with regulations remains a top priority, especially as data breaches continue to rise.

The intersection of AI and cybersecurity also raises important questions about the role of human oversight. While AI can automate many aspects of security and data management, human intervention is still necessary to ensure the ethical use of AI technologies. The case studies reviewed in this paper highlight the importance of maintaining a balance between automation and human oversight, particularly in scenarios where AI decisions could have significant consequences.

In the realm of image processing, the advancements discussed by Zangana and Mustafa (2024) in the use of deep learning for image denoising underscore the importance of integrating traditional and modern approaches. While deep learning models offer superior performance in image enhancement, they also require substantial computational resources and large datasets for training. This raises concerns about accessibility and the potential for disparities in technological adoption, particularly in regions with limited resources.

Moreover, the practical implementations of AI in MANETs, as explored by Al-Sanjary et al. (2018), highlight the potential of AI in improving communication and network reliability. However, the dynamic nature of MANETs presents unique challenges, such as maintaining network stability and security in constantly changing environments. These challenges necessitate continuous research and development to ensure that AI-driven solutions can adapt to the evolving needs of MANETs.

Finally, the discussion must consider the broader implications of integrating AI into various domains. While AI offers numerous benefits, it also raises questions about the displacement of jobs, the concentration of power in the hands of those who control AI technologies, and the potential for AI to exacerbate existing inequalities. Addressing these concerns requires a multidisciplinary approach, involving collaboration between technologists, ethicists, policymakers, and industry leaders.

In conclusion, the integration of AI, cybersecurity, and data management presents both significant opportunities and challenges. The case studies and practical implementations discussed in this paper illustrate the transformative potential of these technologies, while also highlighting the need for careful consideration of the ethical, social, and technical implications of their adoption. Future research should focus on developing strategies to mitigate the risks associated with AI, ensuring that its benefits are accessible to all, and promoting the responsible and equitable use of these technologies across different sectors.

## 9. CONCLUSION

The integration of cybersecurity, artificial intelligence (AI), and data management has become a cornerstone in the development of advanced technological solutions across various industries. This paper has explored the multifaceted interactions between these domains, highlighting the significant benefits they offer when combined. From enhancing security protocols with AI-driven tools to optimizing data management through intelligent systems, the convergence of these technologies has demonstrated its potential to revolutionize the way we approach complex challenges.

The case studies and practical implementations reviewed in this paper have underscored the transformative impact of AI in cybersecurity, particularly in detecting and mitigating threats with unprecedented speed and accuracy. However, these advancements are not without challenges. The ethical, privacy, and security concerns associated with AI adoption, as well as the technical and operational hurdles, must be carefully addressed to ensure that these technologies are used responsibly and effectively.

Moreover, the discussion has emphasized the importance of maintaining a balance between automation and human oversight. While AI can significantly enhance efficiency and decision-making processes, human judgment remains crucial in ensuring the ethical application of AI technologies, particularly in high-stakes scenarios.

The advancements in image processing and mobile ad hoc networks (MANETs) further illustrate the diverse applications of AI and data management across different fields. These innovations, while promising, also highlight the need for continued research and development to overcome the challenges posed by resource constraints, network stability, and the dynamic nature of evolving technological environments.

In conclusion, the integrative approaches in cybersecurity, AI, and data management offer immense potential for driving innovation and improving operational efficiency. However, realizing this potential requires a comprehensive understanding of the challenges and risks associated with these technologies. Future efforts should focus on fostering collaboration between stakeholders, developing robust ethical frameworks, and ensuring that the benefits of these technologies are accessible and equitable. By addressing these considerations, we can harness the power of AI and data management to create more secure, efficient, and resilient systems for the future.